\documentclass[useAMS,usenatbib]{mn2e}
\usepackage{epsfig}
\usepackage{times}
\usepackage{subeqn}

\newcommand{\abs}[1]{\left\vert#1\right\vert}

\def \dash {^{\,\prime}}

\def \grad {\nabla} \def \del {\nabla}
\def \centreline {\centerline}
    
\def \beqn {\begin{equation}}
\def \eeqn {\end{equation}}
\def \bdm {\begin{displaymath}}
\def \edm {\end{displaymath}}

\newcommand{\E}[1]{\times 10^{#1}}
 \def \kpc {\,\mathrm{kpc}}  \def \solarmass {M_{\odot}}
\def \kmpersec {\,\mathrm{km}\,\mathrm{s}^{-1}}

\def \sech{\; \mathrm{sech} \;}

\def \d {\:\mathrm{d}}
 \def \dr {\d r}

\newcommand{\pdfrac}[2]{\frac{\partial #1}{\partial #2}}

\def \half {\frac{1}{2}}

\newcommand{\text}[1] {\; \textrm{#1} \;}

\newcommand{\eqref}[1]{(\ref{#1})}
\newcommand{\tfrac}[2]{\fontsize{8}{9.6} \frac{#1}{#2} }
\newenvironment{remark*}{\textit{Remark.}}{}

\def\spose#1{\hbox to 0pt{#1\hss}}
\def\simlt{\mathrel{\spose{\lower 3pt\hbox{$\mathchar"218$}}
     \raise 2.0pt\hbox{$\mathchar"13C$}}}
\def\simgt{\mathrel{\spose{\lower 3pt\hbox{$\mathchar"218$}}
     \raise 2.0pt\hbox{$\mathchar"13E$}}}

\def \dr {\d r}
\def \rhat {\hat{r}}   
  \def \vecalpha {\balpha}
\def \alphax {\alpha_{x}} \def \alphay {\alpha_{y}}

\def \Ds {D_{\mathrm{s}}} \def \Dl {D_{\mathrm{l}}} \def \Dls {D_{\mathrm{ls}}}

\def \kappap {\kappa_{\mathrm{p}}} \def \gammap {\gamma_{\mathrm{p}}} 
\def \kappabar {\bar{\kappa}}
\def \sigmacrit {\sigma_{\mathrm{cr}}}
\newcommand{\DiagMatrix}[2]{\left(\begin{array}{cc} #1 & 0 \\ 0 & #2 \end{array}\right)}
\def \bysqrt {\sqrt{y^{2} + b^2}} \def \Bzsqrt {\sqrt{B^{2}\!+\!z^{2}}}
\def \md {m_{\mathrm{d}}} \def \mb {m_{\mathrm{b}}} 
\def \rc {r_{\mathrm{c}}} \def \rhoc {\rho_{\mathrm{c}}} \def \rhoh {\rho_{\mathrm{h}}}
\def \Md {M_{\mathrm{d}}}  \def \Mc {M_{\mathrm{c}}} \def \Msun {M_{\odot}} \def \Mb {M_{\mathrm{b}}}
\def \alphah {\alpha_{\mathrm{h}}}  
\def \psib {\psi_{\mathrm{b}}} \def \psid {\psi_{\mathrm{d}}} \def \psih {\psi_{\mathrm{h}}}
\def \psixx {\psi_{xx}}  \def \psixy {\psi_{xy}}  \def \psiyy {\psi_{yy}}
\def \rlambdasqrt {\sqrt{1 + \lambda^{2}r^{2}}}

\title[The Milky Way Galaxy as a Lens]
{The Milky Way Galaxy as a Strong Gravitational Lens}

\author[E.M. Shin and N.W. Evans] {E.M. Shin\thanks{E-mail:
ems@ast.cam.ac.uk; nwe@ast.cam.ac.uk} and
N.W. Evans\footnotemark[1]\\ Institute of Astronomy, University of
Cambridge, Madingley Road, Cambridge, CB3 0HA, United Kingdom}

\begin{document}

\pagerange{\pageref{firstpage}--\pageref{lastpage}} \pubyear{2006}

\maketitle
\label{firstpage}

\begin{abstract}
We study the gravitational lensing effects of spiral galaxies by
taking a model of the Milky Way and computing its lensing
properties. The model is composed of a spherical Hernquist bulge, a
Miyamoto-Nagai disc and an isothermal halo.  As a strong lens, a
spiral galaxy like the Milky Way can give rise to four different
imaging geometries. They are (i) three images on one side of the
galaxy centre (`disc triplets'), (ii) three images with one close to
the centre (`core triplets'), (iii) five images and (iv) seven
images. Neglecting magnification bias, we show that the core triplets,
disc triplets and fivefold imaging are roughly equally likely.  Even
though our models contain edge-on discs, their image multiplicities
are not dominated by disc triplets. The halo is included for
completeness, but it has a small effect on the caustic structure, the
time delays and brightnesses of the images.

The Milky Way model has a maximum disc (i.e., the halo is not
dynamically important in the inner parts). Strong lensing by nearly
edge-on disc galaxies breaks the degeneracy between the relative
contribution of the disc and halo to the overall rotation curve.  If a
spiral galaxy has a sub-maximum disc, then the astroid caustic shrinks
dramatically in size, whilst the radial caustic shrinks more
modestly. This causes changes in the relative likelihood of the image
geometries, specifically (i) core triplets are now $\sim 9/2$ times
more likely than disc triplets, (ii) the cross section for threefold
imaging is reduced by a factor of $\sim 2/3$, whilst (iii) the cross
section for fivefold imaging is reduced by $\sim 1/2$.  Although
multiple imaging is less likely (the cross sections are smaller), the
average total magnification is greater. The time delays are smaller,
as the total projected lensing mass is reduced.
\end{abstract}


\section{Introduction}

In gravitational lensing, galaxies are often represented in a very
idealized manner. For example, models in which either the potential or
the density are stratified on similar concentric ellipses have been
widely studied~\citep{KK93, Ko94, Wi96, Wi97, Hu01, Ev02}. Such
simple, analytic models often yield valuable insights and are
reasonable enough as representations of elliptical galaxies.  They are
less well-suited for spiral galaxies, in which the effects of the
disc, bulge and halo all need to be taken into account.

Gravitational lensing by spiral galaxies has received attention from
\citet{Ke98}, who studied the properties of Mestel and Kuzmin discs
embedded in isothermal haloes. \citet{Wa97} found an analytically
tractable model of a homogeneous disc embedded in an isothermal
sphere, which is nonetheless realistic enough to illustrate a number
of the important effects. The presence of a disc makes lensing effects
sensitive to inclination. In particular, a nearly edge-on disc causes
three images to form on the same side of the lens, a configuration
which has possibly been observed in APM08279+5255
~\citep{Ib99}. \citet{Mo98} used ray-tracing to confirm some of the
theoretical results of \citet{Ke98}.

Of the $\sim 90$ strong lenses, only a handful are known to be spiral
galaxies \citep[see e.g.,][]{Ja00,Win03}. The most extensively studied
spiral galaxy strong lens is B1600+434 \citep[see e.g.,][]{Koo98,
Ja00}. This is a two-image system, composed of a nearly edge-on lens
at $z=0.414$ lensing a background quasar at $z=1.6$. \citet{Win03}
studied another two-image system, PMN J2004-1349, for which the lens
is an isolated spiral galaxy with an inclination angle of $\sim
70^\circ$.  There are at least two known four-image lenses, Q2237+0305
\citep[see e.g.,][]{Schm98,Tr02} and B0712+472 \citep[see
e.g.,][]{Ja98,Ka04}. \citet{Ba98} suggest that the underabundance of
spiral lenses in optical searches may be caused by extinction by dust
and that the missing lenses can be recovered in radio searches.
\citet{Ba00} makes the important point that lensing by spiral galaxies
may dominate over ellipticals once angular resolutions of the order of
$0.1''$ can be achieved. He showed that for the projected Next
Generation Space Telescope, there may be of the order of 10 quasars
per square degree brighter than $V \approx 26$ lensed by spirals.

\citet{Ma00} and \citet{Win03} emphasise the unique contribution that
studies of gravitational lensing of edge-on spirals can provide.  In
the Milky Way galaxy, it has been a matter of debate as to whether the
disc is ``maximum'' or not ~\citep[see e.g.,][]{Se88,Sa97}. A spiral
galaxy has a maximum disc if the inner parts of its rotation are
completely dominated by the disc and bulge, rather than the dark
halo. Although recent evidence strongly supports the idea of a maximum
disc for the Milky Way~\citep{En99,Ha00,Bi01}, the relative
contributions of the disc, bulge and the halo to the flat rotation
curves of external spiral galaxies remains uncertain~\citep[see
e.g.][]{Co99,Pa00}. The issue of whether spiral galaxies typically
possess a maximum disc could be resolved by the study of a sample of
edge-on spirals acting as lenses, as gravitational lensing effects are
sensitive to the relative masses in the rounder halo or flatter disc
components.  The first steps in this direction have been taken by
\citet{Ma00} who concluded that the spiral lens of B1600+434 is not
maximum.

In this paper, we pick a model of the Milky Way galaxy with
bulge, disc and halo that is widely-used in Galactic
astronomy~\citep[see e.g.,][]{Jo95,Di99,He04,Fe06} and ask what its
lensing properties would be, if it were acting as a gravitational lens
at a redshift of $z \approx 0.4$. The model is composed of a Hernquist
(1990) bulge, Miyamoto-Nagai (1975) disc and an isothermal halo,
combined so as to reproduce a flattish rotation curve of amplitude
$\sim 220$ kms$^{-1}$ at the Sun. This model has a maximum disc.  We
first discuss the lensing properties of the Hernquist bulge and
Miyamoto-Nagai disc in \S2 and \S3, before combining them with an
isothermal halo in \S4 to give the full Milky Way model.  We
investigate how the lensing properties change as the masses of the
components are varied -- in particular, as the disc changes from
maximum to sub-maximum.

\begin{figure}
\epsfysize=12cm \centerline{\epsfbox{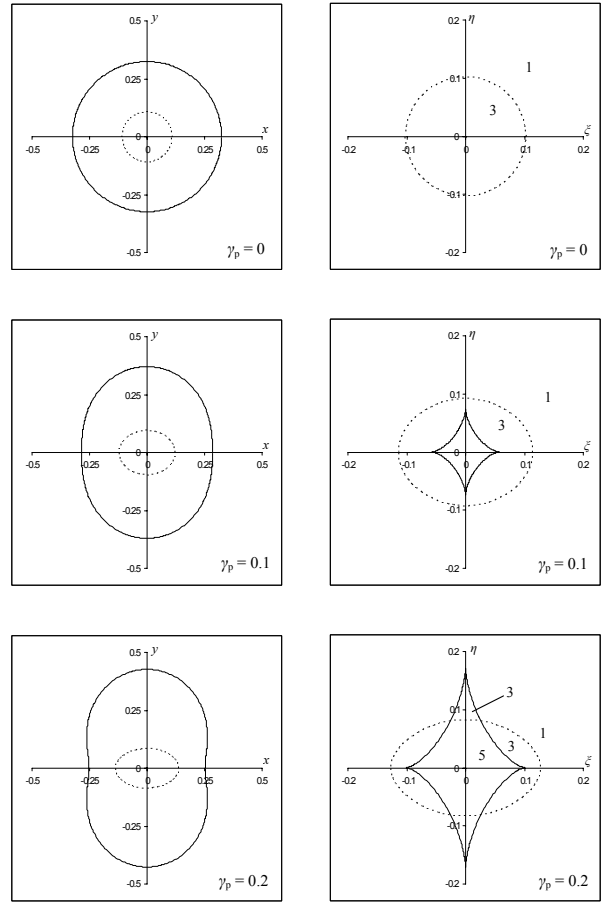}}
\caption{\label{fig:her} Critical curves (left) and caustics (right)
of Hernquist lenses with different strengths of shear. The regions of
multiple imaging in the source plane are labelled. In the top panels
($\gammap=0$), spherical symmetry ensures that the tangential caustic
is a degenerate point at the origin. In the middle panels ($\gammap
=0.1$), the tangential caustic (full curve) is wholly contained within
the radial caustic (dotted curve). In the lower panels ($\gammap =
0.2$) , the cusps of the tangential caustic pierce the radial caustic
(so called ``naked cusps'').}
\end{figure}
\begin{figure}
\epsfxsize=10cm \centerline{\epsfbox{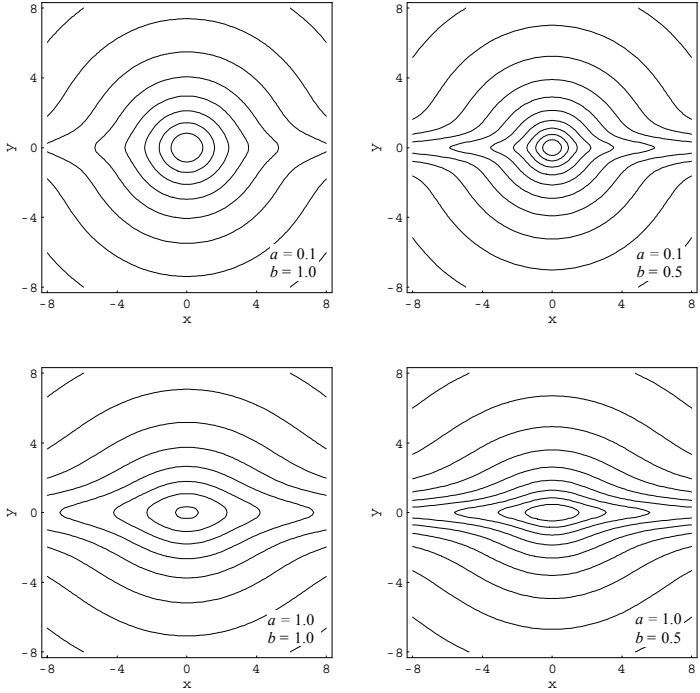}}
\caption{Level curves of the logarithm of the convergence
($\log_{10}\kappa$) for the Miyamoto-Nagai disc with $\md = 1.0.$
Contours are at intervals of 0.5. The projected density can be dimpled
(upper panels) or highly flattened (lower panels).}\label{fig:kapcont}
\end{figure}

\section{The Hernquist Bulge}\label{sec:1-Hernquist}

The Hernquist (1990) model was developed for elliptical galaxies
and the bulges of spiral galaxies.  The three-dimensional Hernquist
mass distribution is
\begin{equation} \label{eq:1-HQ-rho}
\rho(\rhat) = \frac{\Mb}{2\pi}\frac{r_0}{\rhat(\rhat + r_{0})^{3}} \;,
\end{equation}
where $r_{0}$ is a ``core radius'' and $\rhat$ is the spherical polar
radius. Integrating \eqref{eq:1-HQ-rho} along the line of sight
yields
\begin{equation}\label{eq:1-HQ-sigma}
\sigma(R) = \frac{\Mb}{2\pi} \frac{1}{r_{0}^{2}(1-u^{2})^{2}} \left[ (2 + u^{2})\chi(u) - 3 \right] \:,
\end{equation}
where $R$ is the radial coordinate in the lens plane and $u =
R/r_0$, while
\begin{equation}\label{eq:1-Defchi-1}
\chi(u)
=\left\{
\begin{array}
     {l@{\quad}l}
\displaystyle{\frac{\sech^{-1}u}{\sqrt{1-u^{2}}}}\;, & 0 \leq u \leq 1 \;.\\
     \noalign{\vskip7pt}
\displaystyle{\frac{\sec^{-1}u}{\sqrt{u^{2}-1}}}\;, & u \geq 1 \;.\\
\end{array}
\right.
\end{equation}

For convenience, we follow \citet{Sc92} in defining the dimensionless
quantities
\begin{equation}\label{eq:1-DefDimLess-rAndkappaAndsigmacrit}
r = \frac{\rhat}{\xi_{0}}, \qquad \kappa(r) = \frac{\sigma(\xi_{0}
  r)}{\sigmacrit}, \qquad \sigmacrit = \frac{c^{2}\Ds}{4 \pi G \Dl \Dls} \;,
\end{equation}
with $\Ds,\; \Dl,\; \Dls$ being the distances to the source, lens, and
between lens and source, and $\xi_{0}$ an arbitrary scale length. Now,
the bending angle for a circularly symmetric lens is
\begin{equation}\label{eq:1-CircSymalpha}
    \alpha(r) = \frac{2}{r}\int_{0}^{r} r\dash \kappa(r\dash) \dr\dash
    = \tfrac{m(r)}{r}\:.
\end{equation}
Integrating \eqref{eq:1-HQ-sigma} and choosing the scaling $\xi_{0} =
r_{0}$ gives us the bending angle of the Hernquist model as
\begin{equation}\label{eq:1-HQ-alpha-of-r}
    \alpha(r) = \mb\frac{r}{1 - r^{2}} (\chi(r) - 1),\qquad \mb =
    \frac{\Mb}{\pi\sigmacrit r_{0}^{2}}\;.
\end{equation}
Introducing $\kappabar(r) = m(r)/r^{2}$, then the lens
equation including any quadrupole perturbation is
\begin{equation}\label{eq:1-HQ-lenseqn-withGammas}
    \left(\begin{array}{c}
    \xi \\ \eta
    \end{array}\right) = [1 - \kappabar(r)]\left(\begin{array}{c}
                                                x \\
     y\end{array}\right) - \DiagMatrix{\Gamma_{1}}{\Gamma_{2}}
     \left(\begin{array}{c} x \\ y \end{array}\right)\;,
\end{equation}
where $(x,y)$ are dimensionless (length scale $\xi_{0}$) Cartesian
coordinates in the lens plane and $(\xi,\eta)$ are dimensionless
coordinates in the source plane (length scale $\eta_{0} = \xi_{0}
\Ds/\Dl$). The components $\Gamma_{1,2}$ are given by $\kappap \pm
\gammap$, where $\kappap$ is the density of a uniform sheet of
matter superimposed on the lens and $\gammap$ is the shear. Direct
differentiation yields the components of the Jacobian matrix $ A_{ij}
= \pdfrac{\xi_{i}}{x_{j}}$, from which we obtain the image
magnification $\mu = 1 / \det A\:,$
\begin{eqnarray}
A_{11} &=& 1 - \kappabar(r) - \Gamma_{1} -
\frac{x^{2}}{r}\kappabar\dash(r)\;,\nonumber\\
A_{12} &=& A_{21} = -\frac{x y}{r}\kappabar\dash(r) \;, \\
A_{22} &=& 1 - \kappabar(r) -
\Gamma_{2} - \frac{y^{2}}{r}\kappabar\dash(r)\;.\nonumber
\end{eqnarray}
The critical curves are given by $\mu^{-1} \equiv \det A = 0,$ i.e.
\begin{equation}\label{eq:1-HQ-CritCurve-defn}
(1\!-\!\kappabar\!-\!\Gamma_{1}) (1\!-\!\kappabar\!-\!
\Gamma_{2})\!-\!(1\!-\!\kappabar\!-\!\Gamma_{1})\frac{y^{2}}{r}\kappabar\dash\!-\!
(1\!-\!\kappabar\!-\!\Gamma_{2})\frac{x^{2}}{r}\kappabar\dash = 0\:.
\end{equation}
Neither the critical curves nor the image positions can be found
analytically, but both can be found numerically.  The critical curves
and caustics are plotted in Figure~\ref{fig:her} for different
$\gammap$ for $m_{h} = 1, \kappa_{p} = 0$; the numbers of images of
sources in different regions of the source plane are shown. Varying
$m_{h}$ and $\kappa_{p}$ changes the sizes of the critical curves and
caustics but not their asymmetry.  The number of images is odd, not
even. The three-dimensional mass-density is singular at $\rhat=0$ and the
surface density is logarithmically divergent at $r = 0,$ but the
bending angle $\alpha$ and the deflection potential are continuous and
finite, so the odd-number theorem applies. This is consistent with the
study of \citet{Ev98} on cusped density distributions and image
numbers. They found that that cusps less severe than isothermal
($\kappa \propto r^{-1}$) give rise to an odd number of images, but
those more severe than isothermal give rise to an even number of
images.

\begin{figure}
\epsfxsize=6cm \centerline{\epsfbox{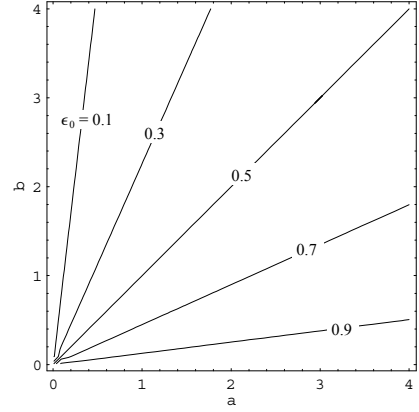}}
\caption{The central ellipticity of the projected mass distribution of
the Miyamoto-Nagai disc is constant along lines of constant gradient
in ($a,b$) parameter space. Along such lines, models closer to the
origin are `diskier' (possess a stronger ridge along the $x$-axis).}
\label{fig:ab-eccentricity-ratio}
\end{figure}
\begin{figure}\epsfysize=12cm \centerline{\epsfbox{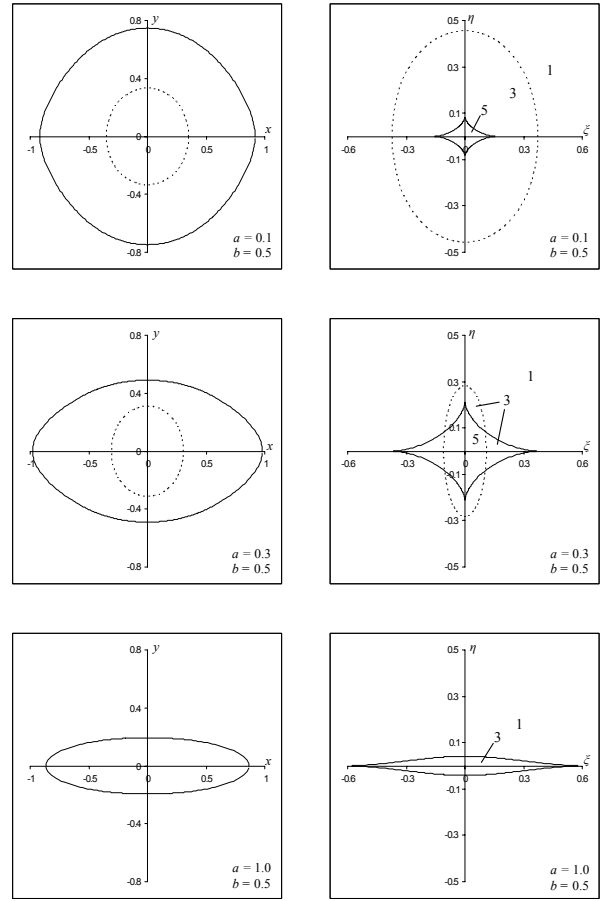}}
\caption{Critical curves (left) and caustics (right) for the
Miyamoto-Nagai disc with $\md = 1.0$ and various values of $a$ and
$b$. The central ellipticiy of the projected density contours is
$0.16$ (upper panels), 0.37 (middle) and 0.67 (lower).  The regions of
multiple imaging in the source plane are labelled. Note that the disc
model contains the same mass, but that the central surface density is
diminishing on moving from the top panels to the bottom panels.}
\label{fig:discimage}
\end{figure}

\section{The Miyamoto-Nagai Disc Model}\label{sec:2-299Disc}

\subsection{Preliminaries}

The Miyamoto-Nagai (1975) disc has the following density distribution:
\begin{equation}\label{eq:3-MN-3D-Density}
\rho(R,z) = B^{2} \Md {AR^{2}\!+\![ A\!+\!3\Bzsqrt]
[A\!+\!\Bzsqrt]^2 \over 4\pi [R^2\!+\!\left(
A\!+\!\Bzsqrt \right)^2 ]^{5/2}[B^{2}\!+\!z^{2}]^{3/2}}\:.
\end{equation}
Here, R and z are cylindrical coordinates. The $A \to 0$ and $B \to 0$
limits are
\begin{equation}\label{eq:2-MN-Case-Ais0}
\rho(R,z) = \frac{3B^{2} \Md}{4\pi}\frac{1}{\left( R^{2} + z^{2} +
B^{2} \right)^{5/2}}\;,
\end{equation}
\begin{equation}\label{eq:2-MN-Case-Bis0}
\rho(R,z) = \frac{A \Md}{2\pi}\frac{1}{\left( R^{2} + A^{2}
\right)^{3/2}} \:\delta(z) \:,
\end{equation}
which are the Plummer (1911) model and the Kuzmin (1956)
disc. The lensing properties of the former are discussed in
\citet{We06} and the latter in \citet{Ke98}.

The Newtonian potential of the Miyamoto-Nagai disc is
\begin{equation}\label{eq:2-MN-GravPot}
\Phi(R,z) = -G M \: \left[R^{2} + \left(A +
\sqrt{z^{2}+B^{2}}\:\right)^{2}\:\right]^{-1/2},
\end{equation}
which is obtained from that of a point mass by the substitution
\begin{equation}\label{eq:2-MN-DefiningSubstitution}
\rhat^{2} \equiv R^{2} + z^{2} \longrightarrow R^{2} + \left(A +
\sqrt{z^{2} + B^{2}} \right)^{2} \:.
\end{equation}

\begin{figure}
\epsfxsize=8cm \centerline{\epsfbox{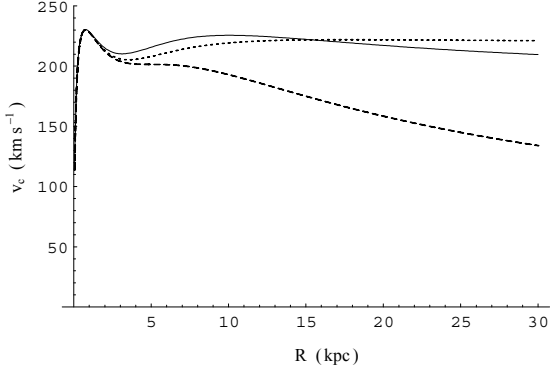}}
\caption{Rotation curve for the model of the Milky Way built with
disc, bulge and halo (solid line). The dashed line shows the
effect of removing the halo component, whilst the dotted line shows
the sub-maximum model.}
\label{fig:rcurve}
\end{figure}
\begin{figure}
\epsfysize=6cm \centreline{\epsfbox{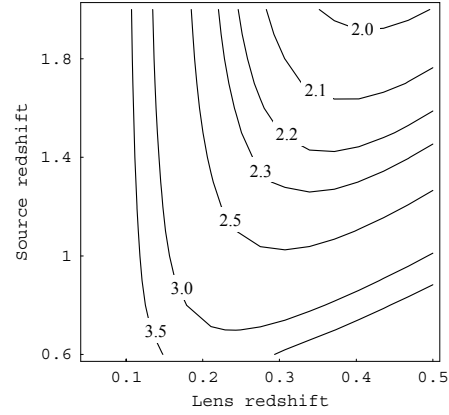}}
\caption{Dependence of $\sigmacrit$ on lens and source redshifts in an
Einstein-de Sitter universe. Contours are in units
$10^9\Msun\kpc^{-2}.$}
\label{fig:eshin}
\end{figure}
\begin{figure*}
\epsfxsize=18cm \centerline{\epsfbox{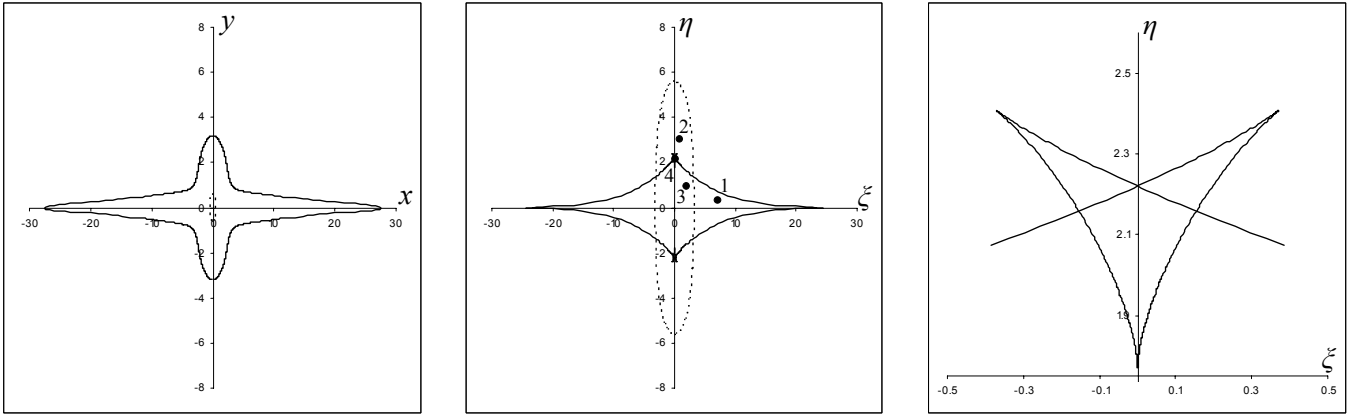}}
\caption{Critical curves (left panel) and caustics (centre panel) for
lensing by a Milky Way galaxy. Note the small dotted critical curve
around the origin. The right panel shows an enlargement of the upper
butterfly cusps in the tangential or `astroid' caustic.  Dots in the
source plane are source positions whose image configurations are shown
in Figure~\ref{fig:fermats}.}
\label{fig:MW-curves}
\end{figure*}

\subsection{The Projected Matter Density}\begin{verbatim}\end{verbatim}

\noindent
The integral of the Newtonian potential along the line of sight (which
is proportional to the deflection potential) and the integral of the
3-D density along the line of sight (which is the surface mass density
relevant to lensing) are not generally analytic. However, in the most
important case when the disc is viewed edge-on, these integrals
\textit{are} analytic. From
\begin{equation}\label{eq:2-MN-EdgeOn-Sigma-as-integral-of-rho}
\Sigma(x',z') = \int_{-\infty}^{+\infty} \rho(x',y',z') \d y'
\end{equation}
we obtain the dimensionless surface mass density
\begin{equation}\label{eq:2-299-kappa}
\kappa = \frac{ b^{2}\md [2\sqrt{b^{2}\!+\!y^{2}}
(2a^{2}\!+\!b^{2}\!+\!y^{2})\!+\!a (a^{2}\!+\!
5b^{2}\!+\!x^{2}\!+\!5y^{2})] }{ 2(b^{2} + y^{2})^{3/2}
[a^{2}\!+\!b^{2}\!+\!x^{2}\!+\!y^{2}\!+\!2a
\sqrt{b^{2}\!+\!y^{2}}]^{2} } \;.
\end{equation}
where we have set
\begin{displaymath}
x = \frac{x'}{\xi_{0}} \:,\quad y = \frac{z'}{\xi_{0}}\;,\quad \md =
\frac{\Md}{\pi \sigmacrit \xi_{0}^{2}} \:,\quad
\end{displaymath}
\begin{equation}\label{eq:3-MN-299-Identifications}
a = \frac{A}{\xi_{0}} \:,\quad b = \frac{B}{\xi_{0}} \:,
\end{equation}
so that $(x,y)$ are dimensionless Cartesian coordinates in the lens
plane.

Some level curves of $\log \kappa$ are shown in
Figure~\ref{fig:kapcont}. The model is useful, as the surface density
distribution can be highly flattened.  We see that $a$ controls the
overall extent of $x$-$y$ asymmetry of the surface mass density ($a =
0$ is circularly symmetric), while $b$ controls the sharpness of the
ridge along the x-axis. Asymptotically, $\kappa$ falls off along the
x- and y-axes as
\begin{subequations}\label{eqns:2-299-Asymptotics}
\begin{equation}
\kappa(x,0) = \frac{\md \, a}{2b} \: x^{-2} + \mathrm{O}(x^{-4}) \;,
\end{equation}
\begin{equation}
\kappa(0,y) = b^{2}\md\: y^{-4} + \mathrm{O}(y^{-5}) \;.
\end{equation}
\end{subequations}
Near the centre, the level curves of $\kappa$ are ellipses with
semiaxes $|a_{20}|$ and $|a_{02}|$, where
\begin{subequations}\label{eqns:2-299-kappa-TaylorCoefficients}
\begin{equation}\label{eq:2-299-kappa-a20}
a_{20} = -\md \; \frac{a + 4b}{2b (a + b)^{4}} \;, \quad\text{and}
\end{equation}
\begin{equation}\label{eq:2-299-kappa-a02}
a_{02} = -\md \; \frac{3a^{2} + 9 a b + 8b^{2}}{4b^{3}(a + b)^{3}} \;
\end{equation}
\end{subequations}
are coefficients in the Taylor expansion about the origin
\begin{equation}\label{eq:2-299-kappa-Taylor}
\kappa(x,y) = \kappa(0,0) + a_{20}x^{2} + a_{02}y^{2} + \mathrm{O}(x^{4},y^{4},x^2y^2)\:.
\end{equation}
The isodensity contours (or isophotes given a constant mass-to-light
ratio) of the model near the origin are the ellipses given by
\eqref{eqns:2-299-kappa-TaylorCoefficients} and
\eqref{eq:2-299-kappa-Taylor}. The central ellipticity is
\begin{equation}\label{eq:2-299-epsilon-near-origin}
\epsilon_0 = 1 - \sqrt{a_{20}\over a_{02}} = 1 - \sqrt{{2(c + 4) \over (c +
1)(3c^{2} + 9c + 8) }}
\end{equation}
where $c \equiv a/b$. This vanishes when $c=0$, as it should. Contours
of constant central ellipticity are straight lines in the ($a,b$)
plane, as shown in Figure~\ref{fig:ab-eccentricity-ratio}.  Moving
along a line of given central ellipticity $\epsilon_0$ away from the
origin takes us from models that are disc-like in the outer parts to
models that are intrinsically rounder.

\subsection{Disc lensing}

\noindent
The deflection potential, related to the surface mass density by
$\kappa(x,y) = \frac{1}{2} \del^2 \psi\:, $ is
\begin{equation}\label{eq:2-299-DeflectionPotential-Dimless}
\psi = \half \: \md \: \log \left[ x^{2} + \left( a + \bysqrt
\right)^{2} \right] \:.
\end{equation}
Note that the $ b \to 0 $ limit is
\begin{equation}\label{eq:2-299-DefPotWhen-bIsZero}
\psi = \half \: \md \: \log \left[ x^{2} + (a + \abs{y})^{2} \right],
\end{equation}
not to be confused with the deflection potential of a point mass at $y
= -a,$ which is identical except $\abs{y} \rightarrow y$. We remark
that the very simple projected potential of the Miyamoto-Nagai disc
seems not to have been noticed before. It is an attractive model for
studies of lensing of arbitrarily flattened mass distributions.

The bending angle $\vecalpha = \grad \psi$ is algebraic, with
components
\begin{eqnarray}
\alphax &=& \frac{\md \: x}{ x^{2} + \left( a + \bysqrt \right)^{2} }\;,
\nonumber \\
\alphay &=& \frac{\md \: y \left( a + \bysqrt
\right)}{\bysqrt \left(x^{2} + \left( a + \bysqrt \right)^2\right)} \;.
\end{eqnarray}
The critical curves and caustics can again be found numerically,
solving
\begin{equation}
\det A \equiv (1 - \psixx) (1 - \psiyy) - (- \psixy)^{2} = 0
\end{equation}
by Newton-Raphson along radial lines.  A sample showing qualitatively
different configurations is shown in Figure~\ref{fig:discimage}, with
the number of images for sources in different regions marked.  In the
upper panel ($\epsilon_0 = 0.16$), the tangential caustic is contained
within the radial caustic. In the middle panel ($\epsilon_0 = 0.37$),
the cusps of the tangential caustic are ``naked''. The density
contours become flatter still in the lower panel ($\epsilon_0 =
0.67$), and the tangential caustic has vanished. This is easily
understood, as although the disc mass is the same, the central density
diminishes on moving from the top to the bottom panels. Hence, this is
the sequence from the two-lips to the single-lips caustic~\citep[see
e.g.,][]{KK93}.

\begin{table}
\begin{center}
\caption{Mean magnifications and time delays (in days) for the core
triplet, disc triplet and five image geometries for the Milky Way
Model, the sub-maximum disc, and the Milky Way model with halo removed
\label{tab:edone}}
\begin{tabular}{cccc}
\hline
\null & Milky & Sub-Maximum & Halo\\
\null & Way   & Disc        & Removed\\
\hline
\textbf{5-image systems} & \null & \null & \null \\
Total magnification & 4.8 & 7.2 & 3.6 \\
$\mu_{5}$ & 2.5 & 3.6 & 1.9 \\
$\mu_{4}$ & 1.8 & 2.7 & 1.4 \\
$\mu_{3}$ & 0.37 & 0.72 & 0.30 \\
$\mu_{2}$ & 0.14 & 0.26 & 0.12 \\
$\mu_{1}$ & 0.005 & 0.007 & 0.006 \\
$t_{5} - t_{1}$ & 39.0 & 23.9 & 35.4 \\
$t_{4} - t_{1}$ & 31.6 & 18.8 & 29.1 \\
$t_{3} - t_{1}$ & 10.8 & 3.9 & 11.4 \\
$t_{2} - t_{1}$ & 8.5 & 3.0 & 8.7 \\
\null & \null & \null & \null \\
\textbf{Core triplets} & \null & \null & \null \\
Total magnification & 2.1 & 2.6 & 1.7 \\
$\mu_{3}$ & 1.6 & 2.0 & 1.2 \\
$\mu_{2}$ & 0.51 & 0.59 & 0.42 \\
$\mu_{1}$ & 0.02 & 0.02 & 0.02 \\
$t_{3} - t_{1}$ & 56.8 & 34.8 & 51.4 \\
$t_{2} - t_{1}$ & 51.3 & 31.5 & 46.6 \\
\null & \null & \null & \null \\
\textbf{Disc triplets} & \null & \null & \null \\
Total magnification & 4.2 & 6.2 & 3.4 \\
$\mu_{3}$ & 1.9 & 2.7 & 1.5 \\
$\mu_{2}$ & 1.6 & 2.3 & 1.3 \\
$\mu_{1}$ & 0.69 & 1.17 & 0.58 \\
$t_{3} - t_{1}$ & 1.27 & 0.30 & 1.54 \\
$t_{2} - t_{1}$ & 1.04 & 0.24 & 1.30 \\
\hline
\end{tabular}
\end{center}
\end{table}
\begin{figure}
\epsfxsize=9.5cm \centerline{\epsfbox{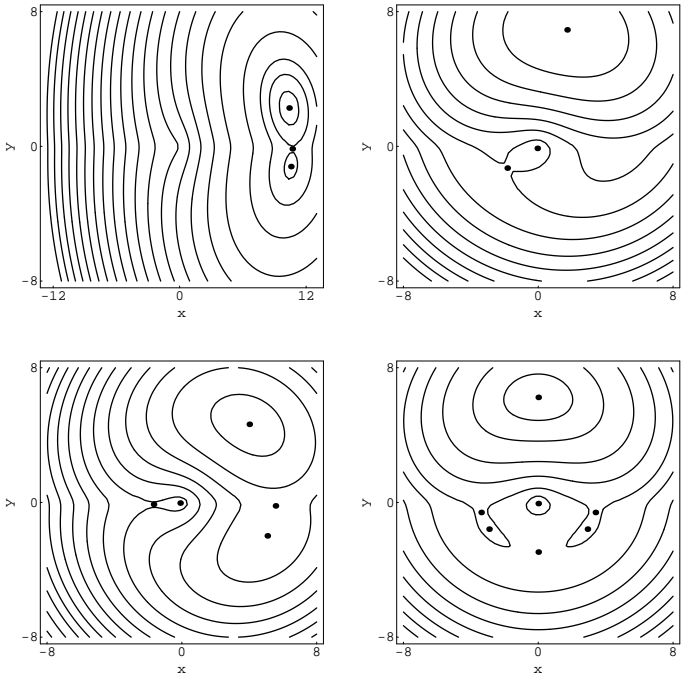}}
\caption{Fermat surfaces and image positions for the sources marked 1
to 4 in Figure \ref{fig:MW-curves}.  Top left: Source 1 produces a
`disc' three-image configuration, with two brighter images straddling
a demagnified one in the plane of the disc.  Top right: Source 2
produces a three-image geometry. The central image is highly
demagnified.  Bottom left: Five-image geometry from source 3. Again,
the images in the plane of the disc are demagnified, with the central
one highly demagnified. The images at (4,4.6) and (5,-2) have
magnifications of about 1.7.  Bottom right: Seven-image geometry of
source 4, which is in a butterfly cusp. The highest magnification
image is at ($0,-2.9$) with a magnification of about $7$; the next
highest are at ($\pm 2.9$,-1.6) with a magnification of about $5.6$.}
\label{fig:fermats}
\end{figure}
\begin{figure}
\epsfxsize=7.5cm \centerline{\epsfbox{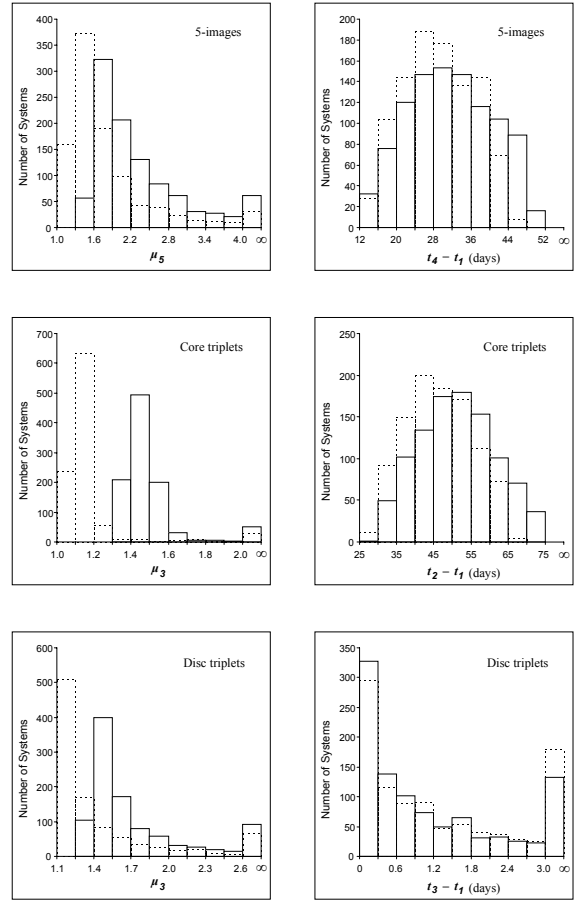}}
\caption{Histograms of the magnification of the brightest image (left
column) and the maximum observable time delay between images (right
column) for the five-image, core triplet and disc triplet
configurations for the Milky Way model. The effect of removing the
halo is indicated by the dotted histograms.}
\label{fig:histos}
\end{figure}
\begin{table} 
\begin{center}
\caption{Cross sections for the three, five and seven image geometries
for the Milky Model, the sub-maximum disc, and the Milky Way model 
with halo removed
\label{tab:edtwo}}
\begin{tabular}{cccc}
\hline 
\null & Milky Way & Sub-Maximum Disc & Halo Removed\\
\hline
7 image & 0.03 & 0.008 & 0.02 \\
5 image & 0.52 & 0.23 & 0.53 \\
\null & \null & \null & \null \\
Total 3 image & 1.57 & 1.03 & 1.61 \\
Core triplet & 0.89 & 0.85 & 0.82 \\
Disc triplet & 0.68 & 0.19 & 0.79 \\
\hline 
\end{tabular}
\end{center}
\end{table}

\section{The Milky Way as a Strong lens}\label{sec:3-MilkyWay}

\subsection{A Model for The Milky Way}\label{subsec:3-MilkyWay}

A widely used model of the Milky Way is a combination of a Hernquist
bulge, a Miyamoto-Nagai disc and a cored isothermal halo ~\citep[see
e.g.,][]{Pa90,Jo95,Di99}. The Newtonian potential (up to an additive
constant) is
\begin{equation}
\Phi = \Phi_{\mathrm{b}} + \Phi_{\mathrm{d}} + \Phi_{\mathrm{h}} \:,
\end{equation}
with $\Phi_{\rm b}$, $\Phi_{\rm d}$, and $\Phi_{\rm h}$ the potentials
of the bulge, disc and halo respectively:
\begin{eqnarray}\label{eq:3-HQ-NewtonianPotential}
\Phi_{\mathrm{b}} &=& \frac{-G \Mb}{\rhat + r_{0}}\;, \nonumber\\
\Phi_{\mathrm{d}} &=& \frac{-G \Md}{\sqrt{R^{2} + \left(A + \sqrt{z^{2}
+ B^{2}}\right)^{2}}} \;, \\
\Phi_{\mathrm{h}} &=& \frac{G \Mc}{\rc} \left[ \half \log(1 +
\frac{\rhat^{2}}{\rc^{2}}) + \frac{\rc}{\rhat}
\tan^{-1}\left(\frac{\rhat}{\rc}\right) \right] \;.\nonumber
\end{eqnarray}
The halo potential corresponds to a cored isothermal sphere
with density
\begin{equation}\label{eq:3-CIS-3D-Density}
\rho(\rhat) = \frac{\rhoc}{1 + \rhat^{2} / \rc^{2}}
\end{equation}
where $\Mc = 4\pi\rhoc\rc^{3}$.  A typical set of vales for the
parameters are $\Mb = 3.4 \times 10^{10} \Msun,$ $r_{0} = 0.7\kpc,$
$\Md = 10^{11} \Msun,$ $A = 6.5\kpc,$ $B = 0.26\kpc,$ $\Mc = 5 \E{10}
\solarmass, $ $\rc = 6.0\kpc$.  The rotation curve of this model is
shown in Figure~\ref{fig:rcurve}. The local circular speed is $\sim
220$ kms$^{-1}$ and the model has a flattish rotation curve out to
$\sim 50$ kpc.

Let us consider lensing by an edge-on galaxy with these
components. (The effect of the disc component decreases rapidly as the
inclination angle departs from zero).  Choosing the scale $\xi_{0} =
r_{0} = 0.7\kpc$, we readily obtain
\begin{equation}\label{eq:3-MilkyWay299-a-and-b}
a \approx 9.3\:,\quad b \approx 0.37 \:, \quad \frac{\md}{\mb} \approx
2.94.
\end{equation}
As for the cored isothermal sphere, its dimensionless projected
density $\kappa (r) = \sigma (r_{0} r) / \sigmacrit$ is
\begin{equation}\label{eq:3-CIS-kappa}
\kappa(r) = \rhoh \left[ 1 + \lambda^{2} r^{2} \right]^{-1/2},
\end{equation}
where the dimensionless $\rhoh$ and $\lambda$ are
\begin{equation}\label{eq:3-Def-rhoh-Def-lambda}
\rhoh = \rhoc \frac{\pi \rc}{\sigmacrit} \:, \qquad \lambda =
\frac{r_{0}}{\rc} \:.
\end{equation}
The bending angle is
\begin{equation}\label{eq:3-CIS-alpha}
\alphah = 2\rhoh \frac{r}{1+\sqrt{1+\lambda^{2}r^{2}}} \:.
\end{equation}
We include a halo for completeness, although we will show later that
it has a small effect on the lensing properties of the model.

\subsection{Typical Image Configurations}

Consider, for concreteness, a galaxy like the Milky Way at a redshift
of 0.4 lensing a quasar at a redshift of 1.5. In an Einstein-de Sitter
universe with Hubble constant of 50 kms$^{-1}$Mpc$^{-1}$, the
redshift-distance relation gives $\sigmacrit \approx 2.17 \times
10^{9} \, \Msun\kpc^{-2},$ and the dimensionless mass/density
parameters are $\mb \approx 10.2\,,\; \md \approx 30\,,\;\rhoh \approx
0.16\,.$ (Figure~\ref{fig:eshin} shows the dependence of $\sigmacrit$
on source and lens redshifts in an Einstein-de Sitter universe. As
distances change, the dimensionless mass/density parameters scale
accordingly). The critical curves and caustics of this lens are shown
in Figure~\ref{fig:MW-curves}.  There are 1, 3 or 5 images for sources
in the different main regions. Notice that there is a tiny seven-image
region within the butterfly cusp, shown enlarged in the rightmost
panel. The formation of butterfly and swallowtail cusps in some lens
models has been noted before -- for example, in the study of disc-like
and boxy ellipticals by \citet{Ev01}, and in the study of the effects
of halo substructure by \citet{Br04}.  It is interesting that a
straightforward disc and bulge system by itself can give rise to such
higher order imaging, a result also found by \citet{Wa97}.

For a fixed source position $(\xi, \eta),$ the images are at
stationary points of the Fermat surface (see e.g. \cite{Sc92})
\begin{equation}\label{eq:3-Def-phi}
\phi_{(\xi,\eta)}(x,y) = \half ( (x - \xi)^{2} + (y - \eta)^{2} ) -
\psi(x,y) \:,
\end{equation}
where $\psi(x,y)$ is the deflection potential of the three-component
(bulge, disc, halo) model:
\begin{equation}
\psi(x,y) = \psib(x,y) + \psid(x,y) + \psih(x,y) \:,
\end{equation}
and
\begin{eqnarray}
\psib &=& \half \mb (\log \frac{r^{2}}{4} + 2\chi(r)),\nonumber \\
\psid &=& \half \md \log \left[ x^{2} + (a + \sqrt{b^{2}+y^{2}})^{2}
\right]\:, \\ \psih &=& 2 \rhoh\lambda^{-2} [ \rlambdasqrt -
\log(1+\rlambdasqrt) ] \nonumber.
\end{eqnarray}
Here, $\psib$ and $\psih$ are obtained from the relation
\begin{equation}
\psi_{j}(r) = 2 \int_{0}^{r} r\dash \kappa_{j}(r\dash) \log(\frac{r}{r\dash}) \: dr\dash.
\end{equation}

In Figure~\ref{fig:fermats} are plotted Fermat surfaces for four
characteristic source positions. The locations of the images are
shown. The multiple-image configurations are different to those of
spherical or elliptical lenses. In all four cases, images near the
plane of the disc, typical of such lenses, are demagnified.  The
threefold image configurations may be split according to the whether
the images are offset to one side of the galaxy centre (`disc
triplets') or whether one image occurs close to the galaxy centre and
the other two on either side (`core triplets'). In fivefold or
sevenfold imaging, there is one central image whilst the remaining
images lie on a broken Einstein ring.

\subsection{Magnifications and Time Delays}

Although interesting, the caustics are not directly observable. In
addition to the positions of images, the observables of any strong
lens may include the time delays and the ratios of image fluxes. The
time delay is related to the Fermat potential via
\begin{equation}
\tau = {\xi_0^2 \over c}{\Ds \over \Dl \Dls} (1+ z_{\rm l}) 
\phi_{(\xi,\eta)}(x,y)
\label{eq:timed}
\end{equation}
where $z_{\rm l}$ is the lens redshift, here taken by $0.4$.  To
investigate the distributions of these quantities, we generate $10^3$
source positions randomly within the caustics for each of the
five-image and three-image cases and record the image properties. The
magnifications are ordered $\mu_1, \mu_2, \dots$ in increasing
order. The means of the least highly magnified image, the second least
highly magnified image, and so on, are recorded in
Table~\ref{tab:edone}, whilst histograms of the distributions are
shown in Figure~\ref{fig:histos}.  Similarly, the time delays are
ordered $t_1, t_2, \dots$ in increasing order. For the 5 image and
core triplet cases, the most delayed image is usually the
unobservable, highly demagnified central image, which corresponds to
$t_5$ or $t_3$ respectively.  The histograms show the maximum
observable time delay, which is the largest time delay between the
remaining images after the central image has been excluded.

Typically, the largest time delays occur for the core triplet
configurations and next for the five-image systems. For lenses
comparable to the Milky Way, it is of the order of month on
average. However, the three images in disc triplets have very similar
arrival times and so the time delays are much smaller, about a day on
average. The images of highest magnification typically occur in
five-image systems, for which the average total magnification is
$4.7$. Disc triplets are more highly magnified than core triplets, the
average total magnifications being $4.2$ and $2.1$
respectively. Notice from Table~\ref{tab:edone} that disc triplets
typically consist of two images of very similar brightness and one
faint image. By contrast, the average magnifications of the three
images of core triplets are more disparate.

\begin{figure}
\epsfxsize=7.cm \epsfbox{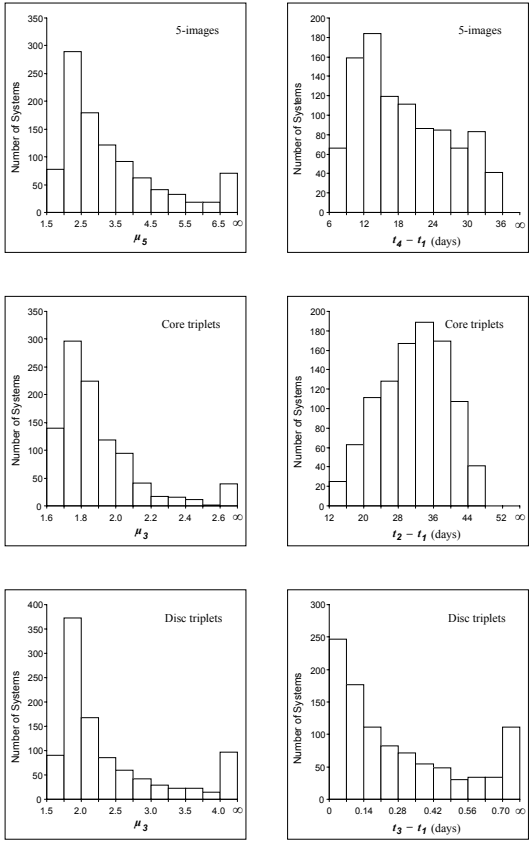}\hspace{1cm}
\caption{As Fig.~\ref{fig:histos}, but for a sub-maximum disc spiral
galaxy.}
\label{fig:histosb}
\end{figure}

\section{Astrophysical Applications}

In this section, we consider changes to our basic Milky Way model to
investigate the lensing characteristics of sub-maximum discs (\S 5.1),
the importance of the dark halo (\S 5.2) and the cross sections to
multiple lensing (\S 5.3).

\subsection{Spiral Galaxies and Maximum Discs}

Here, we construct a spiral galaxy model with a sub-maximum disc. It
has the same bulge as our Milky Way model, but the mass of the disc is
halved. To maintain the amplitude of the rotation curve as $\sim 220$
kms$^{-1}$ between 15 and 30 kpc, the core radius of the dark halo is
adjusted to 4.5 kpc.  In the sub-maximum model, the contributions of
the disc, bulge and halo to the total circular speed $217$ kms$^{-1}$
at the Sun are $110, 123$ and $141$ kms$^{-1}$, respectively. In other
words, the dark halo is now the single largest contributor to the
rotational support at the Solar radius. The rotation curve of the
model is shown in Fig.~\ref{fig:rcurve} as a dotted curve.

Given the caustics, numerical integration gives the areas of the
various regions of multiple-imaging. It is traditional to normalize
the cross sections with that of a singular isothermal sphere with
circular velocity $220\kmpersec$. This cross section is $\pi
(\xi_{\mathrm{sis}}/ \xi_{0})^{2} \eta_{0}^{2},$ where~\citep[see
e.g.,][]{Sc92}
\begin{equation}\label{eq:3-xiSIS}
\xi_{\mathrm{sis}} = \frac{\sigma_{v}^{2}}{\sigmacrit G}\:,
\end{equation}
and the velocity dispersion is related to the circular velocity
$v_{c}$ by $v_{c} = \sqrt{2}\sigma_v$~\citep[see e.g.,][]{Bi87}. Just
as in~\citet{Ke98}, the magnification bias is neglected in the
computation of the cross sections.

The cross sections for the multiple image geometries of the
sub-maximum disc, normalized to that of an isothermal sphere, are
given in Table~\ref{tab:edtwo}. They can be directly compared to the
same quantities for the Milky Way, which are also listed. For the
sub-maximum disc, the cross section of disc triplet imaging decreases
by a factor of $\sim 1/3$, five-imaging decreases by a factor of $\sim
1/2$, whilst the cross section of core triplets remains largely
unchanged. This is because the astroid caustic shrinks dramatically in
size, whilst the radial caustic undergoes a more modest change. In the
absence of selection effects, a clear-cut difference between maximum
and sub-maximum discs is that the former gives rise to core and disc
triplets in roughly equal numbers, while the latter gives rise to
mainly core triplets. 

Table~\ref{tab:edone} shows the magnifications of all the image
configurations, whilst Fig~\ref{fig:histosb} gives histograms of
maximum magnifications and time delays. Compared to the Milky Way, the
total magnifications of all the image geometries has increased. This
is because the size of the caustics has reduced, and so the source is
typically close to a caustic when multiple imaging occurs. Although
multiple imaging is less likely (the cross sections are smaller), the
images are more highly magnified. Also, the time delays are smaller,
as the total projected lensing mass is reduced.

Thus far, we have ignored the magnification bias. However, we can
roughly estimate its effect on the cross-sections using the mean
magnifications listed in Table~\ref{tab:edone}. Suppose, for example,
the luminosity function of the sources has a slope of -2. Then, the
number of five image systems for the sub-maximum disc actually becomes
similar to that for the Milky Way case. Similarly, the cross-sections
of the core triplets and disc triplets in the sub-maximum case must be
boosted by factors of $\sim 1.5$ and $\sim 2.1$ relative to the Milky
Way case to incorporate the effects of magnification bias.  Even so,
core triplets will still predominate strongly over disc triplets.

\subsection{The Role of the Dark Halo}

Here, we take the Milky Way model and remove the dark halo. The
resulting rotation curve is shown in Fig~\ref{fig:rcurve} as a dashed
line. For the case of the Milky Way, the dark halo has an almost
negligible effect on the size and shape of the caustics.  The removal
of the halo causes changes of $\sim 10 \%$ in the cross sections of
the disc and core triplets, as recorded in
Table~\ref{tab:edtwo}. However, the cross sections for total
three-imaging (i.e., both core and disc) and five imaging are
virtually unchanged. In fact, the cross section for multiple imaging
is very slightly increased with the removal of the halo! This is
because the astroid caustic increases in size slightly more than the
radial caustic shrinks.

The effects of the dark halo on the image magnifications and the time
delays are shown as the dotted histograms in Figure~\ref{fig:histos},
whilst the averages are reported in Table~\ref{tab:edtwo}. Although
removal of the halo causes a slight diminuition in the time delays,
the shape of the time delay histograms are virtually unaffected. There
are, however, some small but noticeable changes in the histograms of
the magnifications. The total magnification diminishes by $\sim 30 \%$
on removing the halo.

In our Milky Way model, the contributions of the disc, bulge and halo
to the total circular speed $225$ kms$^{-1}$ at the Sun are $155, 123$
and $106$ kms$^{-1}$, respectively. In other words, the dark halo
makes a modest contribution to the rotation curve and hence the mass
budget within the Solar circle. The typical scales probed by strong
lensing are only of the order of a few kpc. Therefore, strong lensing
by galaxies such as the Milky Way does not probe the dark matter
distribution in the halo very effectively. We conclude that a
Hernquist bulge and Miyamoto-Nagai disc by themselves provide a
realistic and analytically tractable lensing model for galaxies with
maximum discs like the Milky Way.

\subsection{Cross Sections for Multiple-imaging by Spiral Galaxies}

\begin{figure}
\epsfxsize=9.5cm \centerline{\epsfbox{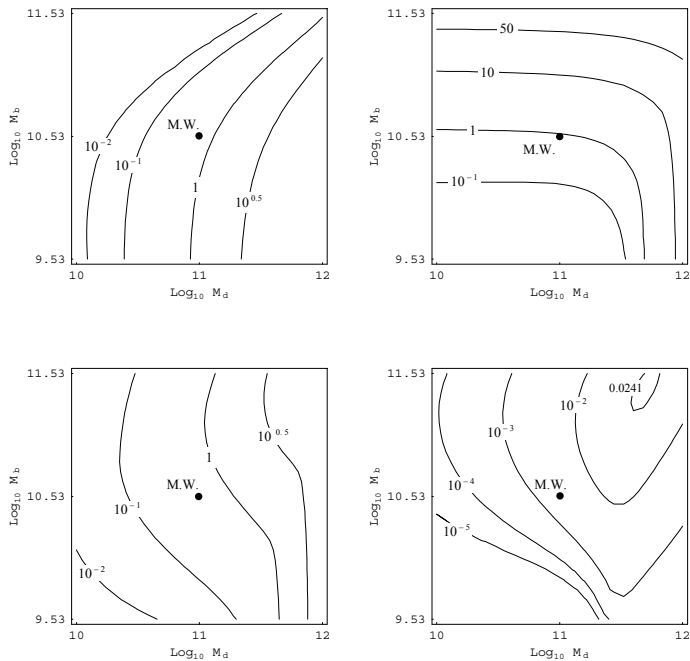}}
\caption{Multiple-imaging cross sections, as a fraction of the
singular isothermal sphere cross section, varying with bulge and disc
masses. Masses are in solar masses (the Milky Way values are in the
centre of each plot). All other parameters are set to Milky Way
values. Disc to bulge mass ratios are constant along diagonals. Top
left: Disc triplet imaging cross section. Top right: Core triplet
imaging cross section. Bottom left: Five-imaging cross section.
Bottom right: Seven-imaging cross section. The butterfly cusp cannot
be made arbitrarily large by making masses larger.}
\label{fig:Xs1}
\end{figure}

\begin{figure}
\epsfxsize=9.cm \centreline{\epsfbox{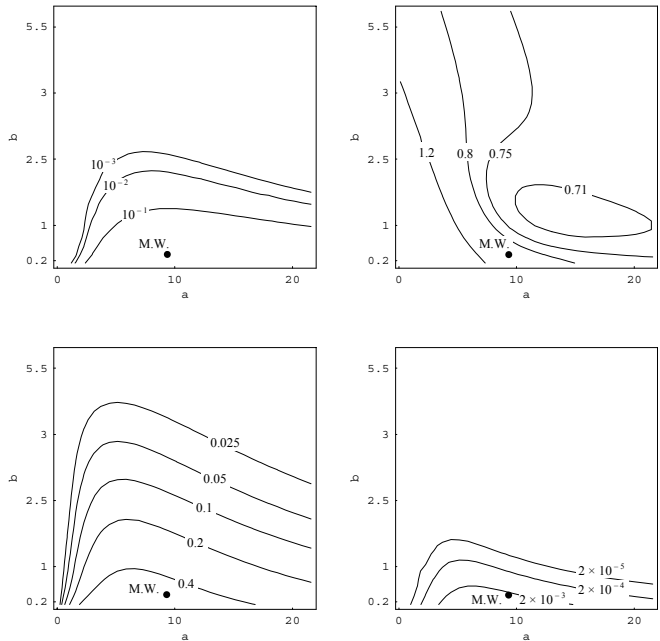}}
\caption{Multiple-imaging cross sections varying with the disc
parameters a and b. All other parameters are set to Milky Way
values. The four panels correspond to the same cross sections as in
Figure~\ref{fig:Xs1}. }
\label{fig:Xs2}
\end{figure}

Figures~\ref{fig:Xs1} and \ref{fig:Xs2} show cross sections of
multiple-imaging varying with the masses of disc and bulge, and with
the shape parameters $a$ and $b$ of the two-dimensional disc model.
The cross sections depend on the distances between lens, source and
observer through $\sigmacrit$, but changing $\sigmacrit$ simply
corresponds to rescaling the dimensionless mass/density parameters
$\mb,$ $\md,$ $\rhoh,$ so cross sections for $\sigmacrit \neq 2.17
\times 10^{9} \Msun \kpc^{-2}$ can also be read off from
Figure~\ref{fig:eshin}, noting also that the singular isothermal
sphere cross section, from \eqref{eq:3-xiSIS}, goes as
$\sigmacrit^{-2}$. In all the panels, the position of the Milky Way
galaxy is shown by a solid circle.

From Figure~\ref{fig:Xs1}, we can read off the cross sections for a
galaxy like the Milky Way (see also Table~\ref{tab:edtwo}). The
cross sections show that the core triplets, disc triplets and fivefold
imaging are roughly equally likely.  Seven imaging configurations are
roughly a thousandfold times less likely. Increasing the mass of the
bulge $\Mb$ at fixed disc mass $\Md$ causes the cross section for disc
triplets to diminish and that for core triplets to increase.
Increasing the mass of the disc $\Md$ at fixed bulge mass $\Mb$ causes
the cross sections for fivefold and sevenfold to increase.

To interpret Figure~\ref{fig:Xs2}, we recall that that the parameter
$a$ controls the ellipticity ($a=0$ is circularly symmetric), while
$b$ controls the sharpness of the disc profile, $b \rightarrow 0$
corresponds to a razor thin disc). Increasing $b$ at fixed $a$ causes
the cross sections of disc triplets and quintuplets to fall sharply,
whilst leaving the cross section for core triplets largely unchanged.


\section{Conclusions}

We have analyzed the strong lensing properties of a realistic model of
the Milky Way galaxy with a disc, bulge and halo, combined to produce
a nearly flat rotation curve. All three components -- the spherical
Hernquist bulge, the Miyamoto-Nagai disc and the cored isothermal halo
-- have analytic deflection angles. There is strong evidence that the
Milky Way galaxy possesses a maximum disc, in the sense that the
contributions to the rotation curve by the disc and bulge at the Solar
radius dominate over the contribution of the halo. The consequence of
this is that the halo has a small effect on the strong lensing
properties, including the magnifications and the time delays.

As a strong lens, the Milky Way galaxy exhibits four different kinds
of multiple imaging geometries. They are (i) three images on one side
of the galaxy centre (`disc triplets'), (ii) three images with one
close to the center (`core triplets'), (iii) five images and (iv)
seven images. For spiral galaxies, the lensing cross sections are
dominated by edge-on models, as this is where the disc makes its
presence known most effectively~\citep{Ke98}. The edge-on case is also
the most important from the point of view of applications.  In this
instance, the cross sections show that the core imaging, disc imaging
and fivefold imaging are roughly equally likely.  The disc triplet
cross section is also sensitive to the thickness of the disc and the
mass of the bulge.

Disc triplets are a characteristic of gravitational lensing by spiral
galaxies with maximum discs.  They occur when the astroid caustic
pierces the radial caustic. They consist of three images straddling
one side of the lensing galaxy. The two outermost images are usually
of comparable magnification, whilst the intervening image is much
fainter.  All three images have very similar arrival times and so the
time delays are small, less than a day on average for our
representative model.  For comparison, the largest time delay between
the observable images in core triplet or fivefold imaging is of the
order of month on average.

Spiral galaxy lenses are rare compared to early-type galaxy lenses.
The importance of gravitational lensing by nearly edge-on spirals is
that it directly probes the shape of the matter distribution and
therefore can break the degeneracy between flat disc and round
halo. Face-on spirals are ineffective in discrimination, as the
projected matter distribution from the disc is round. Even though only
a few examples of nearly edge-on spiral galaxy lenses are known, this
has already provided some important results. For example, \citet{Ma00}
has concluded that the spiral lens of B1600+434 does not have a
maximum disc from detailed modelling of the positions and flux ratios
of the two visible images.

The lensing properties of spiral galaxies with sub-maximum discs
differ from maximum discs.  The astroid caustic shrinks significantly
in size, whilst the radial caustic shrinks more modestly.  Therefore,
the cross section of disc triplets is substantially reduced, whilst
the cross section of core triplets remains roughly unchanged.  The
overall effect is to reduce the cross section to threefold (disc and
core triplets) imaging by $\sim 2/3$ and to fivefold imaging by
$1/2$. The total magnification of the configurations is on average
increased, but the time delays are decreased.

If maximum discs are typical, the lensing cross sections suggest that
core triplets predominate only slightly over disc triplets. However,
if sub-maximum discs are typical, then core triplets are $\sim 9/2$
times more likely than disc triplets.  In fact, except for
APM08279+5255, no disc triplet configurations are known. One possible
interpretation of this scarcity is that the Milky Way is atypical and
that spiral galaxies are usually sub-maximum. This conclusion derives
some support from the detailed modelling of \citet{Ma00} in the case
of B1600+434.
  
It is also interesting that sevenfold imaging can occur for disc
galaxies~\citep[c.f.][]{Wa97,Ev01}, even if the cross section is lower
by a factor $\sim 10^{-3}$ compared to the cross sections for
threefold or fivefold imaging. Our cross sections do not incorporate
the effects of magnification bias, which will boost the likelihood by
a significant amount~\citep{Tu84}.  Note that the existence of higher
order cusps increases the magnification of the quintuple image
configurations as well. This is because the total magnification of the
5 images is increased by the appearance of the butterfly cusps inside
the 5 image caustic.  This fact has been neglected in all calculations
of the numbers of expected 3 or 5 lens systems in the literature.

\section*{acknowledgemnts}
EMS thanks the Commonwealth Scholarship Commission and the Cambridge
Commonwealth Trust for the award of a Studentship. We are grateful to
the anonymous referee for a number of helpful suggestions. This work
was supported by the European Community's Sixth Framework Marie Curie
Research Training Network Programme, Contract No.  MRTN-CT-2004-505183
"ANGLES"



\label{lastpage}

\end{document}